\documentclass[usenatbib]{emulateapj}

\begin{document}

\title{Caustic and Weak Lensing Estimators of Galaxy Cluster Masses}

\author{Antonaldo Diaferio,\altaffilmark{1} 
Margaret J. Geller\altaffilmark{2} and Kenneth J. Rines\altaffilmark{3}}

\begin{abstract}
There are only two methods for estimating the mass distribution in the 
outer regions of galaxy clusters, where virial equilibrium does not hold: 
weak gravitational lensing and identification of caustics in redshift space. 
For the first time, we apply both methods to three clusters: A2390, MS1358 and Cl~0024.  
The two measures are in remarkably good agreement out to $\sim 2 h^{-1}$ Mpc from the 
cluster centers. This result demonstrates that the caustic technique is a valuable complement to
weak lensing. With a few tens of redshifts per $(h^{-1} {\rm Mpc})^2$ within the cluster, the
caustic method is applicable for any $z\lesssim 0.5$.

\end{abstract}


\altaffiltext{1}{Dipartimento di Fisica Generale ``Amedeo Avogadro'', Universit\`a degli Studi di Torino, Via P. Giuria 1, I-10125, Torino, Italy, diaferio@ph.unito.it}
\altaffiltext{2}{Smithsonian Astrophysical Observatory, 60 Garden St., Cambridge, MA 02138, USA, mjg@cfa.harvard.edu}
\altaffiltext{3}{Yale Center for Astronomy and Astrophysics, Yale University, P.O. Box 208121, New Haven, CT 06520-8121, USA, krines@astro.yale.edu}

\section{Introduction}

The relative distributions of mass and light in the universe have remained a
profound and central mystery in cosmology
for more than seventy years. Since Zwicky's pioneering
use of the virial
theorem to discover dark matter in the Coma cluster \citep{zwicky33}, the range and
sophistication of methods for estimating cluster masses and mass profiles
have increased to include a host of dynamical measures, X-ray
estimates and strong and weak gravitational lensing determinations.

Different mass estimators applied to rich clusters of galaxies
constrain the mass distribution on different scales. Strong lensing
generally provides constraints on very small scales ($\lesssim 0.1h^{-1}$ Mpc). 
Virial mass estimates, including Jeans' analysis, assume dynamical equilibrium 
and apply only within the virial radius. Mass estimates based on X-ray
observations assume hydrostatic equilibrium and
rarely extend beyond one-half of the virial radius (\citealt*{maje02}; \citealt{pratt02}).
  
At larger clustrocentric radii where equilibrium assumptions break down,
there exist only two techniques for mass estimation: weak lensing \citep*{kaiser95} and the 
redshift-space caustic technique (\citealt{diaf97}; \citealt{diaf99}, D99 hereafter). 
Both techniques enable determination of the
mass distribution from the cluster center to distances larger
than the virial radius.

The caustic technique has been applied to many local clusters (\citealt{rines03}
and references therein). At small clustrocentric radii, 
caustic estimates agree well with the traditional virial analyses in the optical
and X-ray bands. At larger radii, 
the caustic technique is still valid, but its mass estimates were tested against 
$N$-body simulations only (D99). 

Here we discuss the first comparison of mass estimates from
the caustic technique and weak lensing.  Only recently have sufficient lensing
and spectroscopic data become available to make this comparison.  Both
techniques have known systematic uncertainties: these comparisons test the
importance of these systematics.

In this Letter, we examine mass profile measurements for three intermediate redshift clusters:
A2390, MS1358+6254, and Cl~0024+1654. 

\begin{deluxetable*}{lccccccccccc}
\tabletypesize{\scriptsize}
\tablecolumns{12}
\tablewidth{0pt}
\tablecaption{Cluster parameters}
\tablehead{
  \colhead{cluster} & \colhead{FOV ($\alpha\times\delta$)} & \colhead{$N_{\rm field}\in[z_1,z_2]$} & \colhead{$N$} & \colhead{$\alpha$} & \colhead{$\delta$} & \colhead{$z$} & \colhead{$\langle v^2\rangle^{1/2}$} & \colhead{$R$} & \colhead{$r_s$} & \colhead{$c$} & \colhead{$r_{200}$}
}
\startdata
  A2390 & $43\farcm8\times\phantom{0}7\farcm4$ & $351\in[0.1,0.4]$ & 210 & $21^h 53^m 35\fs53$ & $17\degr 42\arcmin \phantom{0}3\farcs16$ & 0.2284  & 1154 & 0.85 & $0.14\pm 0.17$ & $11\pm 12$ & $1.5\pm 2.4$ \\ 
MS1358 & $20\farcm9\times 21\farcm4$ & $360\in[0.1,0.5]$ & 282 & $13^h 59^m 49\fs65$ & $62\degr 30\arcmin 55\farcs87$ & 0.3289  &  \phantom{0}996 & 0.76 & $0.14\pm 0.09$ & $7.7 \pm 4.3$ & $1.1\pm 0.9$ \\
  Cl~0024 & $20\farcm0\times 24\farcm3$ & $399\in[0.3,0.5]$ & 251 & $\phantom{0}0^h 26^m 35\fs90$ & $17\degr \phantom{0}9\arcmin 41\farcs10$ & 0.3941 & \phantom{0}937 & 0.74 & $0.12\pm 0.11$ & $8.6\pm 7.7$ & $1.0\pm 1.3$ 
\enddata
 \tablecomments{ Col. 2: Field of view (FOV); Col. 3: no. of galaxies in the FOV within the redshift range $[z_1,z_2]$; 
Col. 4: no. of members; 
Cols. 5-7: cluster center coordinates;  
Col. 8: cluster velocity dispersion; Col. 9: cluster size; Cols. 10-12: NFW fit parameters.
Celestial coordinates are J2000. Velocities are in km s$^{-1}$, lengths in $h^{-1}$ Mpc. }
 \label{tab:cl-params}
\end{deluxetable*}
   
\section{The caustic technique}

Cluster galaxies plotted in a redshift space diagram (line-of-sight
velocity $v$  vs. projected distances $R$ from the 
cluster center) distribute in a characteristic
trumpet shape. The boundaries of this trumpet are called 
caustics (\citealt{kaiser87}; \citealt{eniko89}). 
By assuming spherical symmetry and hierarchical clustering for the formation of
the large-scale structure, the caustic mass estimator relates 
the caustic amplitude, the trumpet width in $v$ at each radius $R$, ${\cal A}(R)$, to 
the escape velocity from the gravitational potential
well generated by the cluster. 

The procedure developed by D99 provides an automatic method for locating the
caustics and determining their amplitude. 
First, the procedure arranges all the galaxies in the field in 
a binary tree and finds the cluster members.
The cluster members determine the center of the cluster, 
its one-dimensional velocity
dispersion $\langle v^2\rangle^{1/2}$, and its 
radius $\langle R\rangle$, the mean projected distance of the members from the cluster center. 
Table \ref{tab:cl-params} lists these quantities for the three
clusters.

The procedure next determines 
the threshold $\kappa$ which enters the caustic equation $f_q(R,v)=\kappa$.
Here, $f_q(R,v)$ is the galaxy
density distribution in the redshift diagram, smoothed with 
an adaptive kernel. The parameter $q$ sets the scaling
between the quantities $R$ and $v$.
We choose the parameter $\kappa$ by minimizing the quantity 
$S(\kappa,\langle R\rangle)=\vert\langle v_{\rm esc}^2\rangle_{\kappa,\langle R\rangle}-4\langle v^2\rangle\vert^2$, 
where $\langle v_{\rm esc}^2\rangle_{\kappa,\langle R\rangle}=\int_0^{\langle R\rangle}{\cal A}^2(R)\varphi(R)dR/
\int_0^{\langle R\rangle}\varphi(R)dR$ is the mean caustic amplitude within $\langle R\rangle$ 
and $\varphi(R)=\int f_q(R,v) dv$. 

D99 shows that the three-dimensional cumulative mass profile can now be estimated as 
\begin{equation}
GM(<r)={1\over 2}\int_0^r{\cal A}^2(R)dR\; .
\end{equation}

The error bars on individual data points are proportional to the inverse of the galaxy
number density within the caustics (D99).
This recipe quantifies the uncertainty in the mass estimate which mostly results from 
deviations from spherical symmetry. The recipe was calibrated on $N$-body simulations
\citep{kauffmann99} that generally showed less cleanly defined caustics than in the real Universe.
Therefore, we suspect that these uncertainties are smaller for real clusters than in the
simulations. 
The small scatter ($\lesssim 30\%$) around the equivalence relation between X-ray
and caustic masses \citep{rines03} suggests that the simulations
indeed overestimate the errors in the caustic technique at small radii.
If $30\%$ represents a rough estimate
of the correct caustic mass uncertainty at all radii, the D99 recipe typically overestimates
this uncertainty by a factor of two. 
Nevertheless, because it is the only available prescription for evaluating the error,
we use the conservative
D99 prescription. Comparison of gravitational lensing and caustic measurements for large samples of
clusters in the redhsift range 0.2-0.8 will test the accuracy of this
recipe. 

\begin{figure*}
\centering
\includegraphics[angle=90,scale=.6]{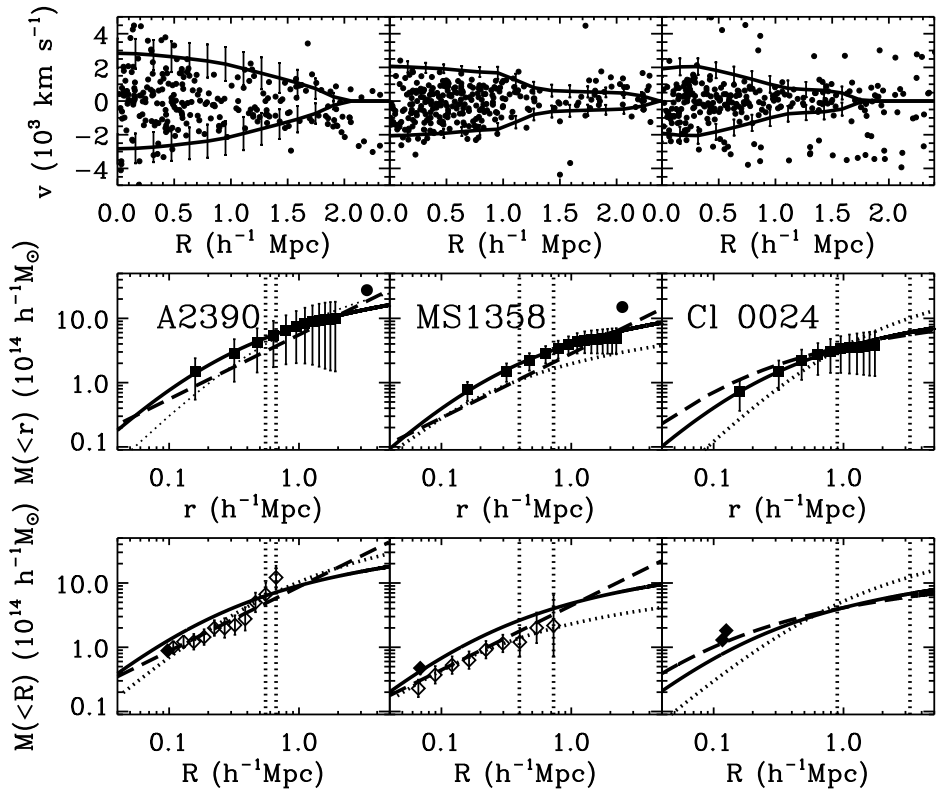}
\caption{The left, middle and right columns are for
A2390, MS1358 and Cl~0024, respectively. {\it Top panels}: Redshift diagrams with the 
galaxies (dots) and caustic locations (solid lines). Line-of-sight velocities $v$ are
in the cluster rest-frame. {\it Middle panels}: Three-dimensional cumulative mass profiles. The solid 
squares show the caustic mass estimates;
the solid lines are the best-fitting NFW profiles to the data points
within $1 h^{-1}$ Mpc; the dotted lines are the best-fitting NFW profiles to the X-ray measures (from left to right: 
\citealt{allen01}; \citealt{araba02}; \citealt{ota04}); the dashed
lines are the best-fitting isothermal (A2390,
\citealt{squires96}; MS1358, \citealt{hoekstra98})  or NFW models  (Cl~0024, \citealt{kneib03}) 
to the gravitational
lensing measures. The left and right vertical dotted lines show the radius of the X-ray
and gravitational lensing fields of view, respectively. 
The two filled circles show the virial estimates by \citet{carlberg96}
of A2390 and MS1358. {\it Bottom panels}: 
Projected cumulative mass profiles; lines are as in the middle panels. 
The open diamonds show the weak lensing measures: A2390, \citet{squires96}; 
MS1358, lower limit by \citet{hoekstra98} to the mass profile.
Filled diamonds show the strong lensing measures: A2390, \citet{pierre96}; 
MS1358: \citet{allen98} from the measurement by \citet{franx97};
Cl~0024: upper symbol, \citet{tyson98}, lower symbol, \citet{broadh00}.
Error bars in all panels are 1-$\sigma$; 
error bars on points where they seem to be missing are smaller than the symbol size. }
\label{fig:comb}
\end{figure*}

\section {Mass comparison }

Mass profile estimates of high-redshift clusters depend on the assumed
cosmological parameters: physical distances, X-ray and weak lensing cumulative mass profiles
scale as the angular diameter distance $D_A$. Moreover, if one derives a best-fitting \citet*{navarro97} (NFW) density 
profile $\rho(r,z) = \delta_c\rho_{\rm crit}(z) (r/r_s)^{-1} (1+r/r_s)^{-2}$,
with $\rho_{\rm crit}(z)=3H^2(z)/8\pi G$ the
critical density of the Universe, $H^2(z)=H_0^2[\Omega_0(1+z)^3 +
(1-\Omega_0-\Omega_\Lambda)(1+z)^2 + \Omega_\Lambda]$, 
$\delta_c = c^3(200/ 3)[ \ln(1+c) -c/(1+c)]^{-1}$
and $c=r_{200}/r_s$ the concentration parameter, $c$ also depends (non-linearly) on $D_A$, 
because $\delta_c\rho_{\rm crit}(z)$ scales as $D_A^{-2}$.
Below, all quantities assume $\Omega_0=0.3$, $\Omega_\Lambda=0.7$
and $H_0=100 h$~km~s$^{-1}$~Mpc$^{-1}$.

Figure \ref{fig:comb} shows the redshift diagrams of the three clusters with the caustic location 
(upper panels) and the mass profiles estimated with the caustic technique, 
gravitational lensing and X-ray data (middle and lower panels).
Gravitational lensing measures all the mass projected onto the sky along the line of sight.
Therefore, we distinguish between three-dimensional (middle panels) and 
projected (lower panels) cumulative mass profiles. Radial distances are three-dimensional
($r$) or projected onto the sky ($R$).

The solid lines in Figure \ref{fig:comb} show the best-fitting NFW profile 
with parameters listed in Table \ref{tab:cl-params}.
To compute these fits, we only used the data points within $r_{\rm lim}= 1 h^{-1}$ Mpc, a
conservative radius beyond which the NFW mass profile might not be a
good description of the actual profile. For all clusters, the data points beyond $1 h^{-1}$ Mpc
do indeed agree with the NFW model, indicating that the correct choice
of $r_{\rm lim}$ is irrelevant.
In any case, the fit parameters and their errors are only indicative, because the individual
data points are correlated. Moreover, the NFW fit parameters are correlated even with independent data points. 
Keeping one of the two 
parameters, $c$ or $r_s$, fixed in our fits reduces their relative errors to $\sim 10$\%.

For each cluster, we also show the best fits determined from the
weak lensing (dashed lines) and X-ray (dotted lines) measurements.
We now comment on each cluster separately.

{\it A2390} is a rich cluster at $z=0.228$ with optical (\citealt{leborgne91};  
\citealt{yee96}) X-ray (\citealt{boehringer98}; 
\citealt*{allen01}) and both weak \citep{squires96} and strong (\citealt{pello91}; \citealt{pierre96}) 
gravitational lensing observations.

\citet{squires96} compare the weak lensing data 
within $\sim 260^{\prime\prime}$ with a singular isothermal model with
velocity dispersion $\sigma=1093$ km s$^{-1}$ taken from \citet{carlberg96}. 
The isothermal model underpredicts the amount of mass actually measured in
the range $0.46-0.67 h^{-1}$ Mpc
(left-bottom panel in Figure \ref{fig:comb}); however, this model
is in good agreement with the best-fitting NFW mass profile derived
by \citet{allen01} from {\it Chandra} observations.
They find $r_s=0.44^{+0.76}_{-0.22} h^{-1}$ Mpc, 
$c=3.6^{+2.0}_{-1.6}$ and $r_{200}=1.6^{+2.9}_{-1.1} h^{-1}$ Mpc.

By using the galaxy redshift survey by \citet{yee96} and by assuming dynamical equilibrium, \citet{carlberg96} estimate 
a mass $M (<3.3 h^{-1} {\rm Mpc}) = (2.7\pm 0.4) \times
10^{15} h^{-1} M_\odot$. The caustic mass $(1.4\pm 1.2) \times 10^{15}h^{-1} M_\odot$ and the 
mass $1.8 \times 10^{15}h^{-1} M_\odot$ extrapolated from the weak lensing
isothermal model are $48\%$ and $33\%$ smaller than this virial mass, 
but within its 3-$\sigma$ uncertainty.

At smaller radii, A2390 sports spectacular 
arcs 
and arclets \citep{pello91}, some of which have measured redshifts (\citealt{beze97}; 
\citealt{frye98}; \citealt{pello99}).
\citet{pierre96} use the brightest strongly lensed arc and its surrounding shear to derive 
the projected total enclosed  mass $M(<97 h^{-1} {\rm kpc}) = (8.0\pm
1.0) {\times}10^{13} h^{-1} M_\odot$\footnote{In this Letter, we rescale each strong lensing
mass found in the literature by the effective lensing distance $D_lD_{ls}/D_s$ 
appropriate to a universe with $\Omega_0=0.3$ and $\Omega_\Lambda=0.7$; $D_l$, 
$D_s$ and $D_{ls}$ are the angular distances to the cluster, 
to the source of the lensed image and between the cluster
and the source, respectively.} (solid diamond in Figure \ref{fig:comb}),
in agreement with the mass $(1.2\pm 0.7)\times 10^{14} h^{-1} M_\odot$ implied
by the projection of the NFW fit to the caustic mass; the strong lensing mass also
agrees with the mass $8.5\times 10^{13} h^{-1} M_\odot$
implied by the weak lensing isothermal model, and is just above the 68\% confidence bound
derived with the X-ray analysis (Figure 8 of \citealt{allen01}).

\citet{pierre96} derive the strong lensing mass by assuming that the
arc is a single lensed galaxy
at $z=0.913$. \citet{frye98} later showed that the fainter part of
this arc actually is a second lensed galaxy at $z=1.033$.
The redshifts of the arcs and arclets,
which are available now but not at the time of Pierre et al.'s analysis,
urges a reformulation of the lensing model of the core of A2390. However, we
expect that a newly derived mass will not substantially differ from the mass
of \citet{pierre96}, because the mass estimated with the simplest
lensing models, which provide the most inaccurate measures, probably
are within 30\% of the true value \citep*{koch03}.

{\it MS1358+6254} is a very rich cluster first discovered by \citet{zwicky68}.
We collect 381 redshifts in the cluster
region from the surveys of \citet*{fabricant91}, \citet{fisher98} and \citet{yee98}.

\citet{hoekstra98} used HST observations to construct a weak
lensing map of the cluster extending to a radius of $\sim 220^{\prime\prime}= 0.73 h^{-1}$ Mpc. 
They only derive a lower limit to the mass profile and find a best-fitting 
singular isothermal model with $\sigma = 780\pm 50$ km s$^{-1}$ 
(dashed lines in Figure \ref{fig:comb}). 
More recently, \citet*{araba02} analyze a {\it Chandra} observation of the cluster.
They approximate the mass profile within $\sim 2^\prime =0.4 h^{-1}$ Mpc with
an NFW profile, with $r_s=88^{+92}_{-47} h^{-1}$ kpc,
$c=9.3^{+3.8}_{-2.5}$ and $r_{200}=0.81^{+0.92}_{-0.49} h^{-1}$ Mpc. 

\citet{carlberg96} assume virial equilibrium to estimate 
 $M(<2.5 h^{-1} {\rm Mpc})=(1.5\pm 0.2)\times 10^{15} h^{-1} M_\odot$ from 
their galaxy redshift survey. This mass is more than 3-$\sigma$ above the weak lensing
isothermal extrapolation $7.0\times 10^{14} h^{-1} M_\odot$ which agrees with the caustic
estimate $(6.5\pm 2.8)\times 10^{14} h^{-1} M_\odot$.  The
extrapolation of the X-ray fit 
yields $3.0\times 10^{14} h^{-1} M_\odot$, a factor of two smaller
than the caustic mass and a factor of five below the virial mass. 
Probably, the assumption of virial equilibrium at this large
distance is unrealistic and the extrapolation of the X-ray profile, 
limited to radii $<0.4 h^{-1}$ Mpc, is unreliable.

In the very central region, \citet{allen98} 
uses the strong lensing observations by \citet{franx97} to derive a projected mass
$M(< 69 h^{-1} {\rm kpc}) = 4.4\times 10^{13} h^{-1} M_\odot$ with a
$20\%$ uncertainty. The projected NFW profile derived from the caustics
yields a perfectly consistent mass $(4.2\pm 1.3) \times 10^{13} h^{-1} M_\odot$.
The projected profiles derived by \citet{hoekstra98} and \citet{araba02} imply
the somewhat lower masses $3.0\times 10^{13} h^{-1} M_\odot$ and $2.9\times 10^{13} h^{-1} M_\odot$,
respectively. 

The X-ray and weak lensing mass models agree within $\sim 0.8 h^{-1}$ Mpc,
but underestimate the strong lensing mass derived by \cite{allen98}.
The fact that the weak lensing mass provides only a lower limit to 
the mass profile and the caustic mass
is in excellent agreement with the strong lensing measurement suggests
that the caustic mass provides the correct mass profile 
of MS1358 out to $\sim 2h^{-1}$ Mpc. 

Significant tension exists between
lensing (\citealt*{bonnet94}; \citealt*{tyson98}) and X-ray  (\citealt{soucail00}; \citealt{ota04}) 
mass estimates of the {\it Cl~0024} cluster.
\citet{kneib03} combine their weak lensing measurements from wide field imaging 
with the strong lensing measurement by \citet{broadh00} to derive the best-fitting NFW profile 
with $r_s=54\pm 2 h^{-1}$ kpc,
$c=18.7^{+7.7}_{-4.3}$ and $r_{200}=1.01^{+0.41}_{-0.23} h^{-1}$ Mpc.
According to their Figure 12, the uncertainty in their mass estimate is always $\lesssim 10\%$.
Our Figure \ref{fig:comb} also shows the NFW profile which fits recent {\it Chandra} data \citep{ota04}. 
These authors derive the NFW profile from a $\beta$-model fit. According to its parameters, we
find $r_s=0.56\pm 0.02 h^{-1}$ Mpc, $c=1.8\pm 0.3$ and $r_{200}=1.02\pm 0.18 h^{-1}$ Mpc.
Our caustic estimate lies between the lensing and the X-ray fits at $r<0.2 h^{-1}$ Mpc, but
it is in excellent agreement with the lensing estimate outside $\sim 0.5 h^{-1}$ Mpc.

In the cluster central region, there are two strong lensing measurements which yield comparable masses.
However, the very small errors claimed make them inconsistent with each other:
$M(<0.114 h^{-1} {\rm Mpc}) = (1.30\pm 0.04)\times 10^{14} h^{-1} M_\odot$ \citep{broadh00}, 
and $M(<0.119 h^{-1} {\rm Mpc}) = (1.563\pm 0.002)\times 10^{14} h^{-1} M_\odot$ \citep{tyson98}.
We scaled the mass reported by \citet{tyson98} by assuming $z=1.675$ for the
arc, as measured by \citet{broadh00}.
By construction, the NFW profile of \citet{kneib03} 
agrees with the former (it yields $M(<0.114 h^{-1} {\rm Mpc}) = 1.13\times 10^{14} h^{-1} M_\odot$) 
and therefore disagrees with the latter (it yields $M(<0.119 h^{-1} {\rm Mpc}) = 1.17\times 10^{14} h^{-1} M_\odot$).
The caustic profile gives smaller, but consistent, masses in both cases: $(7.9\pm 3.8)\times 10^{13}h^{-1} M_\odot$ 
and $(8.5\pm 4.0)\times 10^{13}h^{-1} M_\odot$, respectively. The NFW fit to the X-ray data 
yields even smaller masses: $3.8\times 10^{13}h^{-1} M_\odot$ and $4.2\times 10^{13}h^{-1} M_\odot$
with a $\sim 30\%$ typical error.
\citet{czoske02} suggest that the peculiar redshift distribution of the galaxies within $\sim 3.5h^{-1}$ Mpc 
of the cluster center 
can be explained by a high-speed collision along the line of sight between Cl~0024 and a less massive
cluster. This model implies that the X-ray mass estimate based on dynamical equilibrium
is unreliable. Because the caustic and lensing mass estimators are both independent of the dynamical state of
the cluster, it is reasonable that they agree with each other but disagree with
the X-ray mass.

\section {Conclusion}

For the first time, we compare the only two cluster mass estimators that do not
rely on the dynamical equilibrium of the system: weak gravitational lensing and caustics in
redshift space. We estimate the caustic mass of A2390, MS1358 and Cl~0024 within  $\sim 2h^{-1}$ Mpc
of the cluster center. The caustic mass profiles are in very good agreement with the lensing profiles.
We confirm that the discrepancy between lensing and X-ray mass in Cl~0024 
is probably a consequence of the unrelaxed state of the cluster which invalidates  
the X-ray analysis.

Weak lensing requires accurate photometric wide-field surveys in excellent seeing;
moreover, the cluster sample is somewhat limited to clusters at distances where the
lensing signal is sufficiently strong. Weak lensing 
measures all of the mass projected along the line of sight, resulting in
a minimum 20\% uncertainty in the cluster mass estimates \citep{deputter05}.
The caustic technique, which requires dense wide-field redshift
surveys, provides a complementary measurement of 
the three-dimensional mass profile of individual clusters at moderate redshift;
it also yields robust mass profiles for clusters in the local universe.

Future comparison of these techniques for large samples of clusters,
covering a range of redshifts, will constrain systematic uncertainties in the
methods and may provide insight into the change in the relative amounts of
mass in the infall regions and cluster cores as a function of lookback time.

We thank the referee for noticing a few inaccuracies in the first version
of this Letter. We have made use of NASA's Astrophysics Data System 
and the NASA/IPAC Extragalactic Database (NED) operated by the Jet Propulsion
Laboratory, California Institute of Technology, under contract with NASA.

\end{document}